\documentclass[aps,prb,twocolumn,showpacs,10pt]{revtex4}

\usepackage{graphicx}
\usepackage{times}
\usepackage{color}
\usepackage{amssymb}
\usepackage{amsmath}
\usepackage{amsfonts}

\begin{document}

\title{Spin Liquid Ground State of the Spin-$\frac{1}{2}$ Square $J_1$-$J_2$ Heisenberg Model}

\author{Hong-Chen Jiang}
\affiliation{Kavli Institute for Theoretical Physics, University of
California, Santa Barbara, CA 93106} \affiliation{Center for Quantum
Information, IIIS, Tsinghua
  University, Beijing, 100084, China}
\author{Hong Yao}
\affiliation{Department of Physics, Stanford University, Stanford,
CA 94305, USA} \affiliation{Institute for Advanced Study, Tsinghua
University,
  Beijing, 100084, China}

\author{Leon Balents}
\affiliation{Kavli Institute for Theoretical Physics, University of
California, Santa Barbara, CA 93106}

\date{\today}

\begin{abstract}
  We perform highly accurate density matrix renormalization group (DMRG)
  simulations to investigate the ground state properties of the
  spin-$\frac{1}{2}$ antiferromagnetic square lattice Heisenberg
  $J_1$-$J_2$ model.  Based on studies of numerous long cylinders with
  circumferences of up to 14 lattice spacings, we obtain strong evidence
  for a topological quantum spin liquid state in the region $0.41\leq
  J_2/J_1\leq 0.62$, separating conventional N\'eel and striped
  antiferromagnetic states for smaller and larger $J_2/J_1$,
  respectively.  The quantum spin liquid is characterized numerically by
  the absence of magnetic or valence bond solid order, and non-zero
  singlet and triplet energy gaps.  Furthermore, we positively identify
  its topological nature by measuring a non-zero topological
  entanglement entropy $\gamma=0.70\pm 0.02$, extremely close to
  $\gamma=\ln(2) \approx 0.69$ (expected for a $Z_2$ quantum spin
  liquid) and a non-trivial finite size dimerization effect depending
  upon the parity of the circumference of the cylinder.  We also point
  out that a valence bond solid, and indeed any discrete symmetry
  breaking state, would be expected to show a constant correction to the
  entanglement entropy of {\sl opposite} sign to the topological
  entanglement entropy.
\end{abstract}


\maketitle

\section{Introduction}
\label{sec:introduction}

Quantum spin liquids (QSLs) are elusive magnets without magnetism,
resisting symmetry breaking even at zero temperature due to strong
quantum fluctuations and geometric frustration \cite{Balents2010}.
The simplest QSLs known theoretically are characterized by
topological order\cite{Wen1989, Wen1990I, Kivelson1987}, and support
fractionalized excitations including spinons, which carry the spin
(1/2) but not the charge of the electron. Since the QSL state was
suggested by Anderson\cite{Anderson1973}, it has been sought, mostly
unsuccessfully, in models and materials.  However, exciting
indications of QSL ground states were recently reported in numerical
studies of models on the honeycomb\cite{Meng2010Hubbard} and
kagome\cite{White2011Kagome} lattices. Here we report strong
evidence for a QSL state in the square lattice $J_1$-$J_2$
antiferromagnetic (AFM) Heisenberg model, with the Hamiltonian
\begin{eqnarray}
H &=&J_1\sum_{\langle ij\rangle}\textbf{S}_i\cdot
\textbf{S}_j+J_2\sum_{\langle\langle
ij\rangle\rangle}\textbf{S}_i\cdot
\textbf{S}_j,\label{Eq:ModelHamiltonian}
\end{eqnarray}
where $\textbf{S}_i$ is the spin-$1/2$ operator on site $i$ and
$\langle ij\rangle$ ($\langle\langle ij\rangle\rangle$) denotes
nearest neighbors (next nearest neighbors).  In the following we set
$J_1=1$ as the unit of energy, and consider only the frustrated case
$J_2>0$.

Eq.~(\ref{Eq:ModelHamiltonian}) is of fundamental interest for its
simplicity, and for its relevance to cuprates, Fe-based
superconductors\cite{Seo2008,Si2008,Fang2008,Xu2008,Jiang2009}, and
other materials\cite{Melzi2001}. Accordingly, it is among the most
studied models in frustrated quantum magnetism\cite{Kotov1999,
Dagotto1989, Capriotti2000, Mambrini2006, Wen1991Z, Figurdido1990,
Singh1989, Read1989, Larkin1990,
Zhitomirsky1996,PhysRevB.78.214415,Richter2010}.  These previous
studies have established the existence of a non-magnetic ground
state between the N\'eel and striped AFM states which occur for
small and large $J_2$, respectively.

To characterize the non-magnetic phase, we can ask two main types of
questions.  First, we may ask about its symmetries.  Being
non-magnetic, the ground state retains the internal SU(2)
spin-rotation invariance, but it may break spatial ones.  If SU(2) is
preserved but spatial symmetries are broken in such a way that the
unit cell is enlarged, the system is said to have valence bond solid
(VBS) order.  Second, we may ask about the range of entanglement of
the wavefunction.  The simplest representative wavefunctions for VBS
states are continuously deformable by local unitary transformations
into product states.  Such is true for typical ground state
wavefunctions for systems with broken discrete symmetries (the space
group of a lattice is discrete).  As such, these wavefunctions have
only short-range entanglement (Schr\"odinger cat states are possible
in finite systems and will be discussed in
Sec.~\ref{sec:topol-entang-entr}).  Wavefunctions which {\sl cannot}
be continuously transformed in this way into product states may be
said to exhibit {\sl long-range entanglement}.  This is true for all
gapless critical phases, as well as for some gapped states.  In
particular, gapped QSL states exhibit a particularly simple type of
long-range entanglement, characterized by {\sl Topological
  Entanglement Entropy} (TEE)\cite{Kitaev2006,Levin2006}.  Often the
two types of characterization are conflated, but this is not
necessarily the case. States with both long range entanglement,
e.g. with TEE, {\sl and} VBS order exist. Such states, while not
technically QSLs by the standard definition given above, have all the
same exotic physics as QSLs with unbroken spatial symmetry.  We note,
however, that it is believed that for S=1/2 spins on a lattice such as
this one with an odd number of spins per unit cell, the {\sl absence}
of VBS order {\sl implies} the presence of long-range entanglement.
Therefore a convincing demonstration of vanishing VBS order does,
indirectly, imply interesting QSL physics.  It is, however, less
important to characterizing and proving the existence of a QSL than
positive, direct evidence of long-range entanglement.
\begin{figure}
\centerline{
    \includegraphics[height=2.0in,width=3.4in] {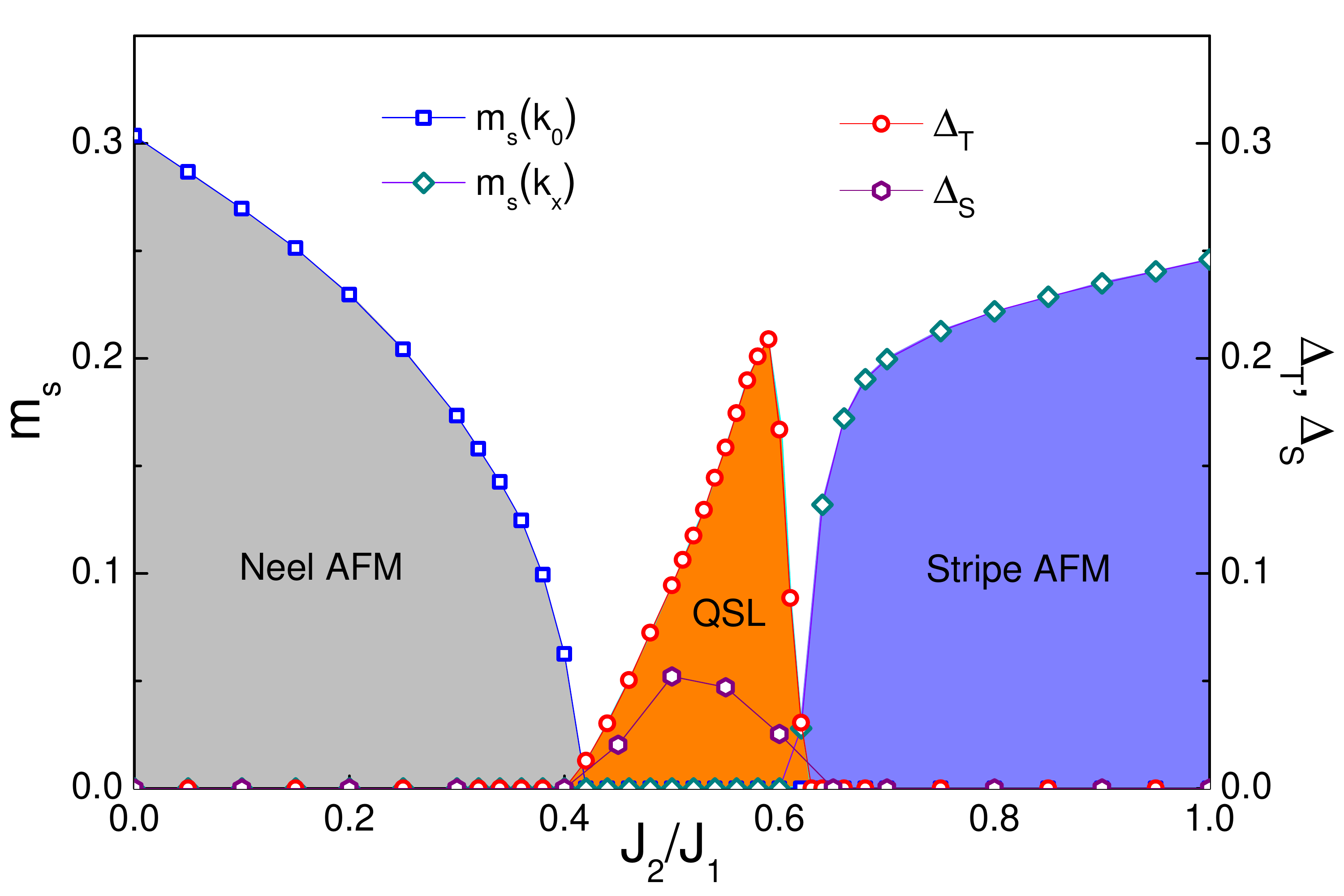}
    }
\caption{(Color online) The ground state phase diagram for the
spin-$\frac{1}{2}$ AFM Heisenberg $J_1$-$J_2$ model on the square
lattice, as determined by accurate DMRG calculations on long
cylinders with $L_y$ up to 14. Changing the coupling parameter
$J_2/J_1$, three different phases are found: N$\rm\acute{e}$el
antiferromagnet (AFM), topological quantum spin liquid (QSL), and
stripe AFM phase. $m_s(\textbf{k}_0=(\pi,\pi))$
[$m_s(\textbf{k}_x=(\pi,0))$] denotes the staggered magnetization in
the N$\rm\acute{e}$el AFM phase [stripe AFM phase], whose saturation
value is $1/2$. $\Delta_S$ and $\Delta_T$ denote the spin singlet
gap and spin triplet gap, respectively.} \label{Fig:PhaseDiagram}
\end{figure}

Most of the literature on the intermediate phase of the $J_1$-$J_2$
model has focused on the possibility of symmetry breaking VBS order.
Many of these prior studied have suggested that the intermediate
state has VBS order.  We note, however, that
all numerical results for the $J_1$-$J_2$ model are based either on
biased techniques (such as series expansion or coupled cluster
methods, or fixed node or related versions of Monte Carlo adapted to
avoid the sign problem which is present for unbiased Monte Carlo in
this system), or on exact diagonalization of very small systems.
Some {\sl
  theoretical} motivation for the possibility of VBS order comes from
the theory of deconfined quantum criticality\cite{deconfinedQCP},
which predicts that a continuous quantum phase transition -- a
deconfined quantum critical point (DQCP) -- should occur between an
ordered Ne\'el state and a plaquette or columnar VBS state, {\sl in
some models}.  However, the existence of such a transition does not
in any way imply that it occurs for the $J_1$-$J_2$ model in
question, or that this particular model even harbors a VBS phase.
Other theoretical motiviation for VBS order comes from its presence
in some large-$N$ generalizations of the nearest-neighbor Heisenberg
antiferromagnet.  However, these large $N$ studies are not
controllably close to the SU(2) case and moreover do not consider
second neighbor interactions. In short, we believe there is very
little compelling evidence for the existence of VBS order in the
isotropic $S=1/2$ $J_1$-$J_2$ model to be found in the prior
literature.  We will return to discuss VBS states in
Sec.~\ref{sec:boundary-effects}.

The {\sl only} unbiased technique capable of treating generic
frustrated two dimensional spin systems of moderately large size is
the Density Matrix Renormalization Group (DMRG)
method.\cite{White1992,White2007,White2011Kagome,Stoudenmire2011}
While the sizes that can be studied using the DMRG are not as large
as those accessibly by quantum Monte Carlo (QMC) for {\sl
unfrustrated} models, they are still very large and they are not
limited by the sign problem, which prevents application of QMC to
most realistic physical models.  Moreover, the DMRG has some
advantages over QMC: it is intrinsically a zero temperature
technique, and obtains a convenient representation of the ground
state wavefunction.  Most importantly for our purposes, the DMRG is
very efficient and convenient for calculating the entanglement
entropy, which we return to in some detail below.  In this paper, we
report the results of extensive simulations (with truncation error
$\sim 10^{-7}$) on numerous cylinders of circumference $L_y=3-14$,
and lengths $L_x\geq 2L_y$.  In our simulations, we measure
spin-spin correlation functions, correlation functions and
expectation values of VBS order parameters, bulk singlet and triplet
energy gaps, and entanglement entropy.  All results confirm the
existence of magnetic order for small and large $J_2$, and that (see
Fig.~\ref{Fig:PhaseDiagram}) the ground state for $0.41\leq
J_2/J_1\leq 0.62$ is non-magnetic, in very good agreement with the
most accurate prior results from series expansion and coupled
cluster\cite{PhysRevB.78.214415} methods.  Furthermore, we find that
the intermediate phase has a gap to both singlet and triplet
excitations and, within our uncertainty, {\sl no} VBS order in the
2D limit as extrapolated from the VBS correlation functions.   We
carry out further checks for possible finite-size effects due to the
boundaries, to see if this might artificially suppress VBS order,
and see no indication that this is the case.

The latter results suggests a QSL state, based on negative evidence:
the apparent absence of VBS order.  We find two {\sl positive}
evidences that this suggestion is correct, and that the state is a
$Z_2$ QSL. First, we find a non-zero TEE, $\gamma$, which is a
constant and universal {\sl reduction} of the von Neumann
entanglement entropy, known to vanish in any gapped state with
short-range entanglement.  Notably, we point out in
Sec.~\ref{sec:topol-entang-entr} that discrete spontaneous symmetry
breaking phases such as valence bond solids have absolute ground
states which are Schr\"odinger cat states with a constant {\sl
enhancement} of the entanglement entropy -- i.e. an effect of {\sl
opposite sign} to the TEE.  Phases with non-zero $\gamma$ and a gap
to all excitations are {\sl topological phases}.  Like conformal
field theories in two dimensions, only discrete types of topological
phases exist, with discrete allowed values of $\gamma$ (which plays
a role somewhat similar to the central charge in a conformal field
theory). For all points we have studied within the non-magnetic
phase, the value of $\gamma$ is equal, within numerical uncertainty
of $2\%$, to $\ln(2)$, which is the {\sl minimal} value possible for
$\gamma$ in a topological phase with time-reversal symmetry.  A
topological entanglement entropy of $\gamma=\ln(2)$ implies either a
$Z_2$ QSL or a ``doubled semion''
phase\cite{jiang12:_ident_topol_order_entan_entrop}.  As there is,
to our knowledge, no theory suggesting the appearance of the semion
phase in an SU(2) invariant spin-$1/2$ model, we take this as strong
evidence for a $Z_2$ QSL state.  The second positive evidence for a
$Z_2$ QSL is a remarkable odd/even effect, in which static VBS order
is entirely absent for even $L_y$ but is observed directly in the
VBS expectation values for odd $L_y$.  This is expected on general
theoretical grounds for a $Z_2$ QSL, as we show in
Appendix~\ref{sec:stagg-dimer}. We compare the behavior of the
numerically observed static VBS order for odd circumference
cylinders with theory, and find quite consistent results.

The remainder of the paper is organized as follows.  In
Sec.~\ref{sec:corr-funct}, we report results of magnetic and dimer
correlation functions, and their extrapolation to the infinite
system limit.  Sec.~\ref{sec:energy-gaps} discusses the singlet and
triplet energy gaps.  Sec.~\ref{sec:topol-entang-entr} describes the
theory and measurements of the topological entanglement entropy, and
Sec.~\ref{sec:odd-even-effect} presents results on the even-odd
effect.  We conclude in Sec.~\ref{sec:discussion} with a summary of
the conclusions, and a detailed discussion of the reasons to think
VBS order, even weak, is unlikely in this model, in response to a
recent critique.\cite{PhysRevB.85.134407}\ The Appendix gives a
theoretical derivation and discussion of some properties of $Z_2$
quantum spin liquids.

\section{Correlation functions}
\label{sec:corr-funct}

In this section we discuss the behavior of correlation functions of
spin and dimer (VBS) operators.  Here and in the rest of the paper,
all our numerical data is based on DMRG simulations on cylinders,
i.e. finite square lattices with $N=L_x\times L_y$ sites and with
open and periodic boundary conditions in the $x$ and $y$ directions,
respectively.  When not otherwise specified, we fix the aspect ratio
to $L_x/L_y=2$, with $L_y=L$, then $L_x=2L$, which has been shown to
optimize results in the
DMRG\cite{White2007,White2011Kagome,Stoudenmire2011}.  Moreover, to
extract bulk properties, we will often work on the central half of
the system with an effective system size $N_c=L\times L$. For
instance, in computing spin correlation functions
$\langle\mathbf{S}_i\cdot \mathbf{S}_j\rangle$, we restrict site
indices $i$ and $j$ to the central half of the system so that the
obtained correlation functions could represent the bulk properties.
We keep more than $m=12000$ states in each DMRG block for most
systems, which is found to give excellent convergence with
truncation errors of the order or less than $10^{-7}$.

\begin{figure}
\centerline{
    \includegraphics[height=7.3in,width=3.2in] {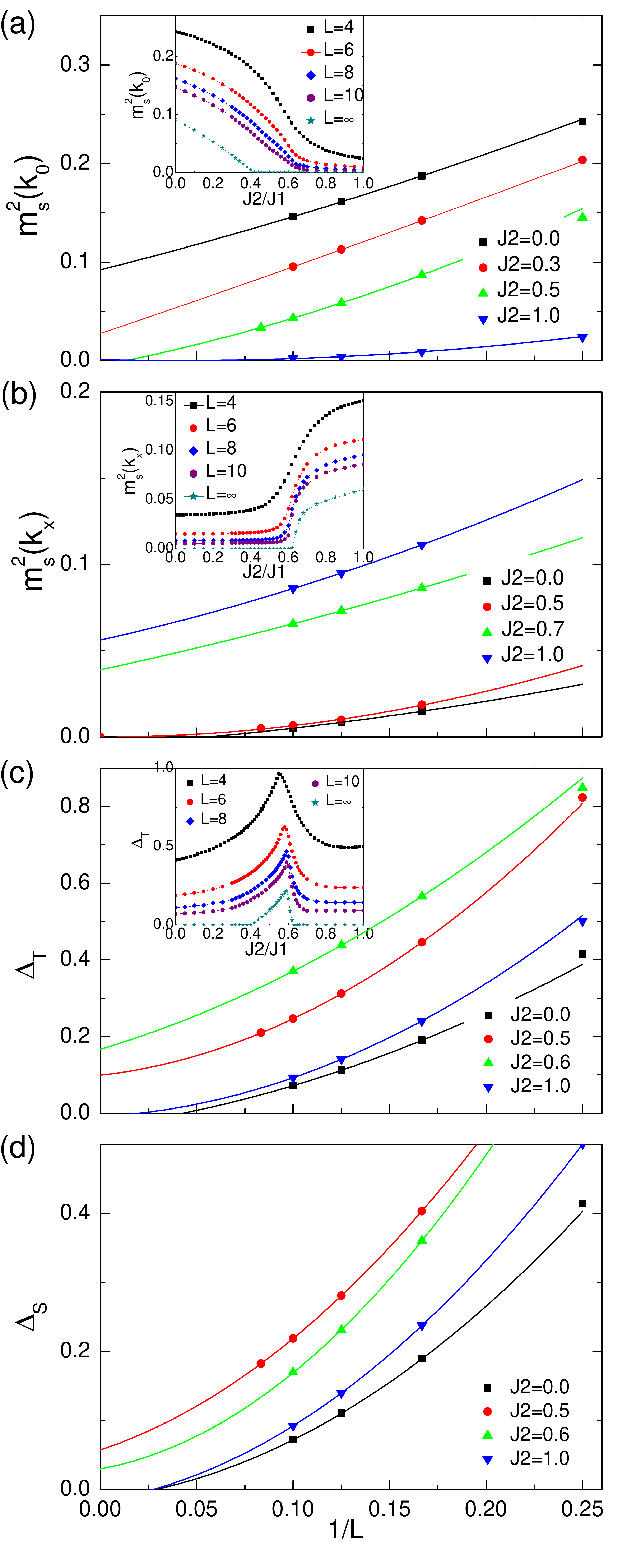}
   }
    \caption{(Color online)
    Finite-size extrapolations of the magnetic
      order parameters and spin excitation gaps. (a) The
      N$\rm\acute{e}$el AFM order parameter $m^2_s(\textbf{k})$ at
      wavevector $\textbf{k}_0=(\pi,\pi)$ and (b) stripe AFM order
      parameter $m^2_s(\textbf{k})$ at wavevector
      $\textbf{k}_x=(\pi,0)$ or $\textbf{k}_y=(0,\pi)$, for various values of
      $J_2$, fitted using second-order polynomials in
      $1/L$. N$\rm\acute{e}$el AFM order disappears for $J_2>0.41$,
      while stripe AFM order develops for $J_2>0.62$, as seen in the
      corresponding insets. (c) Spin triplet gap $\Delta_T$ and (d)
      spin singlet gap $\Delta_S$ for different values of $J_2$, also
      fitted using second-order polynomials in $1/L$. The inset in (c)
      shows $\Delta_T$ for $L=4,6,8,10$, and the extrapolated values
      in the 2D limit, as functions of $J_2$. For the spin singlet
      gap, due to the numerical cost, we
      focus on several typical data points as shown in (d).} \label{Fig:StrFacSpinGap}
\end{figure}

We begin with measurements of the magnetic correlations in the
ground state, $\langle \textbf{S}_i\cdot\textbf{S}_j\rangle$, and
the corresponding static structure factor
$M_s(\textbf{k},L)=\frac{1}{L^2}\sum_{ij}e^{i\textbf{k}\cdot(\textbf{r}_i-\textbf{r}_j)}\langle
\textbf{S}_i\cdot\textbf{S}_j\rangle$.  The structure factor is
peaked at $\textbf{k}_0=(\pi,\pi)$ for small $J_2$ and
$\textbf{k}_x=(\pi,0)$ or $\textbf{k}_y=(0,\pi)$ for large $J_2$,
corresponding to the N\'eel and striped AFM states, respectively. To
quantitatively analyze the order, we perform an extrapolation of the
(squared) staggered magnetization,
$m^2_s(\textbf{k},L)=\frac{1}{L^2}M_s(\textbf{k},L)$, to the two
dimensional limit ($L=\infty$) according to the generally accepted
form
$m^2_s(\textbf{k},L)=m^2_s(\textbf{k},\infty)+\frac{a}{L}+\frac{b}{L^2}$
(see Figs.\ref{Fig:StrFacSpinGap}(a) and (b)).

Extrapolation from data for $L\leq 12$ shows that the N\'eel AFM
order is non-zero for $J_2 <0.41$, while striped AFM order onsets
for $J_2>0.62$, thus establishing the phase boundaries shown in
Fig.~\ref{Fig:PhaseDiagram}.  A strong check on the quality of our
results is the staggered magnetization at $J_2=0$, which we find to
be $m_s(\textbf{k}_0,\infty)$=0.304, very close to the best known
numerical value of the magnetic moment $m_s$=0.307 by large-scale
quantum Monte-Carlo (QMC) simulation\cite{Sandvik1997}.  The
location of the phase boundaries is consistent with previous
studies\cite{Murg2009,Richter2010}.

\begin{figure}
\includegraphics[width=3.2in]{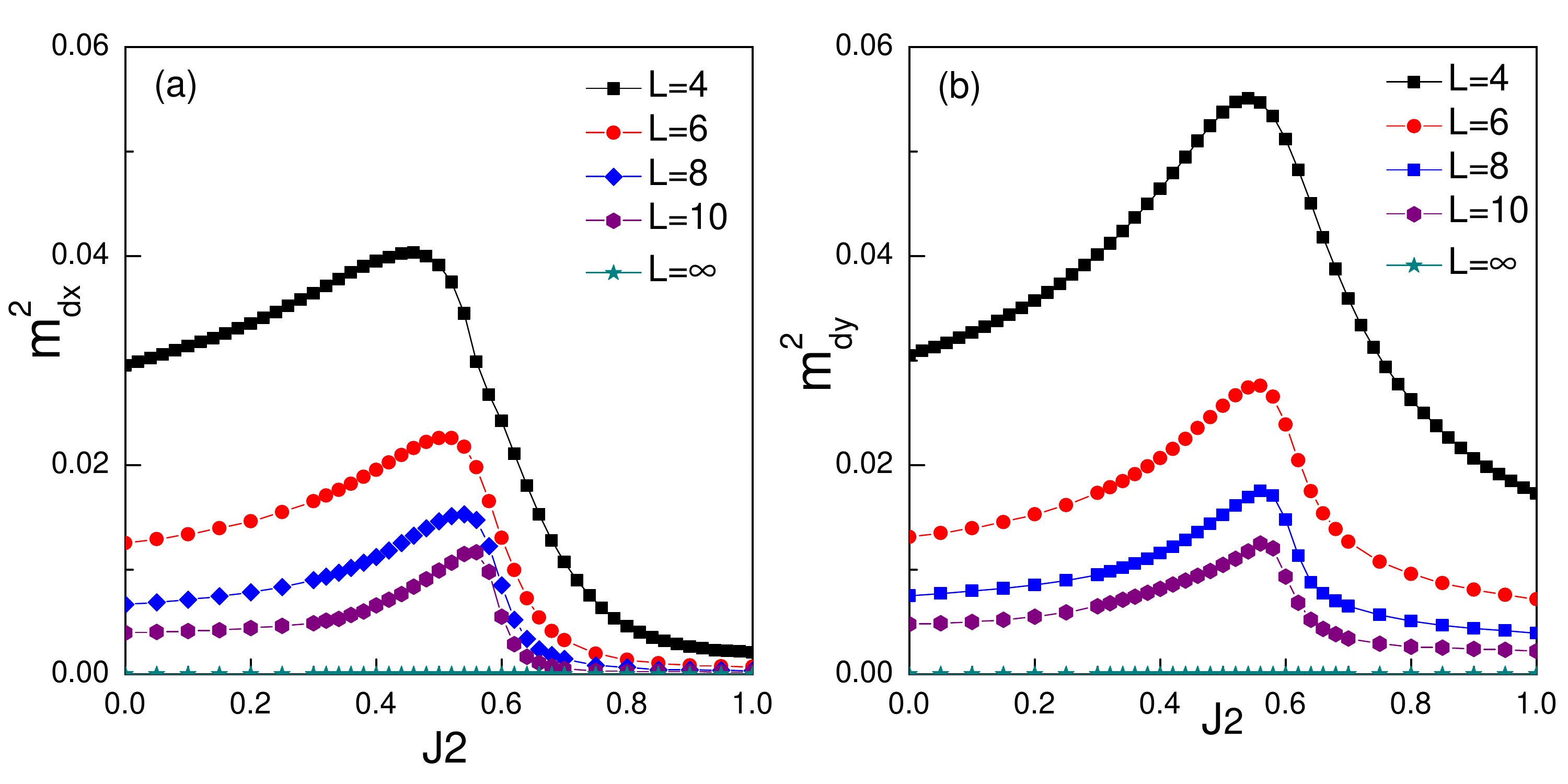}
\caption{(Color online) Order parameters for horizontal and vertical
  dimers.  (a) The dimer order parameter $m^2_{d,x}$ at wavevector
  $\textbf{k}_x=(\pi,0)$ and (b) plaquette order parameter $m^2_{d,y}$
  at $\textbf{k}_y=(0,\pi)$, as a function of $J_2$ for different
  system sizes, and extrapolated to $L=\infty$.}
\label{Fig:DimerDxDy}
\end{figure}

We next consider possible VBS order, which has been considered a
prime candidate for non-magnetic symmetry breaking in the
intermediate phase.  From the bond operators $B^\alpha_i \equiv
\textbf{S}_i\cdot \textbf{S}_{i+\alpha}$ on bond $(i,i+\alpha)$ with
$\alpha=\hat x$ or $\hat y$, we define the dimer-dimer correlation
functions $\langle B^\alpha_i B^\beta_j\rangle$, with the
corresponding structure factor
$M^{\alpha\beta}_{d}(\textbf{k},L)=\frac{1}{L^2}{\sum_{ij}}e^{i\textbf{k}\cdot(\textbf{r}_i-\textbf{r}_j)}
\left(\langle B^\alpha_i B^\beta_j\rangle - \langle
  B^\alpha_i\rangle\langle B^\beta_j\rangle\right)$.  Typical VBS
patterns expected theoretically have momentum $\textbf{k}_x=(\pi,0)$
or $\textbf{k}_y=(0,\pi)$, so to study the correlations, we focus on
$L_y$ even, for which $k_y=\pi$ is an allowed momentum.  We indeed
observe a maximum in $M^{aa}_d(\textbf{k},L)$ at
$\textbf{k}=\textbf{k}_a$ ($a=x,y$), and therefore define the dimer
order parameters by
$m^2_{d,a}(L)=\frac{1}{L^2}M^{aa}_d(\textbf{k}_a,L)$. As shown in
the inset of Fig.\ref{Fig:DimerDxDy}, for finite systems, both
horizontal and vertical dimer order parameters have a maximum within
the intermediate phase.  Note that for the larger systems, the order
parameters for horizontal and vertical dimers become nearly
indistinguishable, indicating that the isotropy of the two
dimensional limit is being recovered.

\begin{figure}
\includegraphics[width=3.2in]{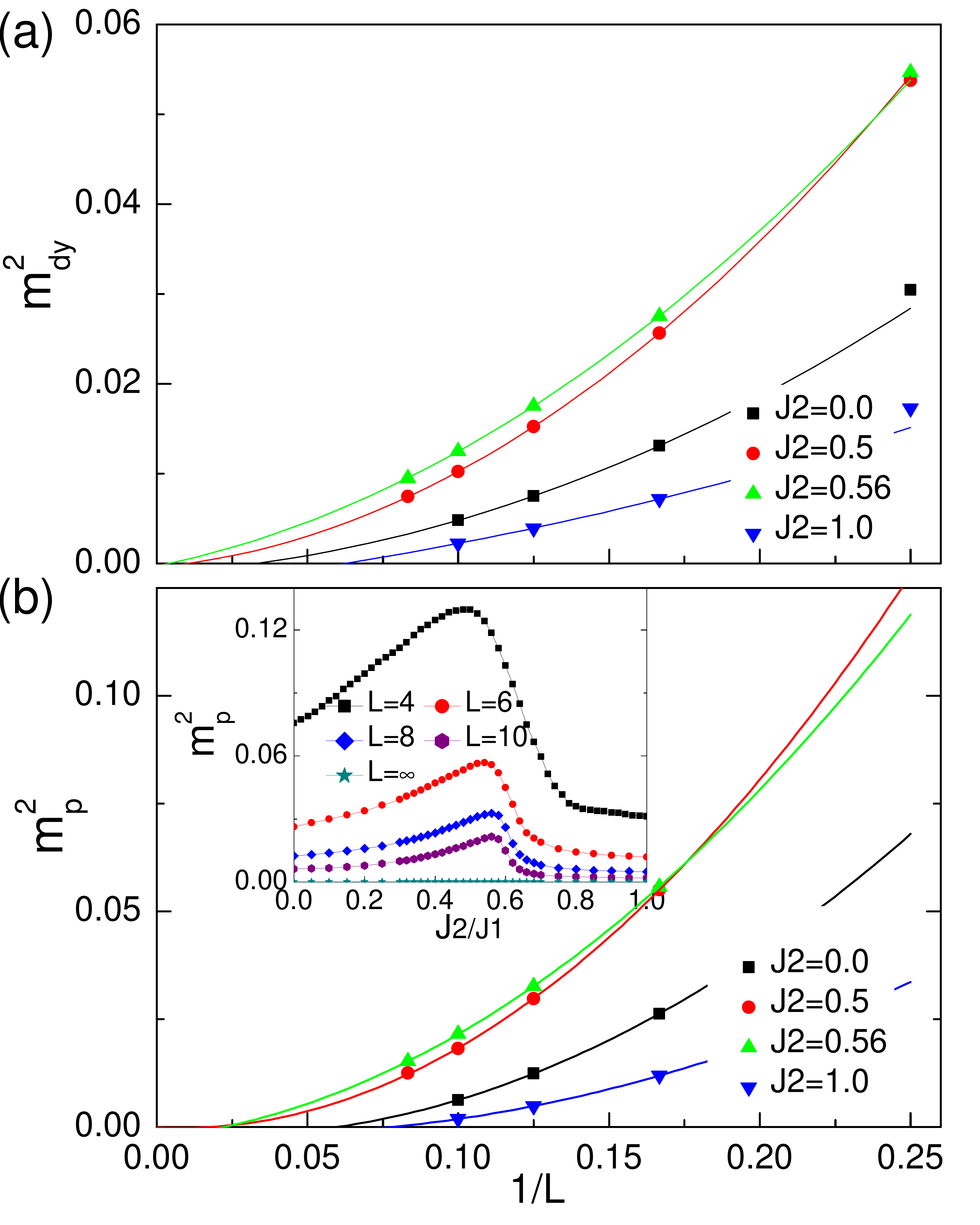}
\caption{(Color online) Finite-size extrapolations of the dimer
order
  parameter and plaquette order parameter. (a) The dimer order
  parameter $m^2_{d,y}$ at wavevector $\textbf{k}_y=(0,\pi)$ and (b)
  plaquette order parameter $m^2_p$, for various values of $J_2$,
  fitted using second-order polynomials in $1/L$. The inset shows the
  plaquette order parameter for $L=4,6,8,10$, and the extrapolated
  values in the 2D limit, as functions of $J_2$.}
\label{Fig:DimerPlaquette}
\end{figure}

Applying the same extrapolation scheme used for the magnetic order
parameters, however, the extrapolated dimerization $m^2_{d,a}$ (see
Fig.\ref{Fig:DimerPlaquette}(a)) for $L\rightarrow\infty$ vanishes
for all $0\leq J_2 \leq 1$.  For characteristic values of $J_2$ near
the middle of the intermediate phase, an exponential fit of the
dimer-dimer correlation function (not shown) gives an estimate of
the VBS correlation length $\xi_d \approx 4$.  Taken at face value,
these observations indicate that the VBS order is a finite-size
effect, and vanishes in the thermodynamic limit.  More
conservatively, at a minimum, the result indicates that the VBS
correlations we observe cannot be distinguished from just {\sl
fluctuation} effects in a state with unbroken spatial symmetry, and
there is no a priori reason to regard them as evidence of true VBS
order.

Both columnar and plaquette VBS phases have been suggested in the
past.   The complex order parameter $m_{d,x}+im_{d,y}$ in fact is
sufficient to detect and distinguish both columnar and plaquette VBS
phases\cite{deconfinedQCP}, but as an additional check we measure
directly the correlations of the plaquette operator
$P_i=\frac{1}{2}(\Pi_i+\Pi^{-1}_i)$ where $\Pi_i$ cyclically
permutes the four spins of the plaquette $i$ in a clockwise fashion.
The plaquette order parameter determined from the corresponding
structure factor (see Supplementary Information) is shown in Fig.
\ref{Fig:DimerPlaquette}(b).  Like the VBS order parameter, it
vanishes in the extrapolation to the thermodynamic limit.

\section{Energy gaps}
\label{sec:energy-gaps}

We next consider the energy gap to bulk singlet and triplet excited
states, and find both to be non-zero in the intermediate phase.
This rules out {\sl any} type of magnetic order, not just the
$(\pi,\pi)$ and $(\pi,0)$ orders considered explicitly via the
correlation functions.  It also rules out other exotic states
breaking SU(2) symmetry, such as spin nematics.  This is because any
state with broken spin-rotational symmetry must have a vanishing gap
by Goldstone's theorem.

To obtain {\sl bulk} excited states, we follow Refs.
\onlinecite{White2007, White2011Kagome, Stoudenmire2011}, and first
target only one state, sweeping enough to obtain a high-accuracy
ground state; then we restrict the range of bonds that are updated
in the DMRG sweeps to the central half of the sample and target the
two lowest-energy states, again sweeping to high accuracy, but
keeping the end regions of the samples locally in the ground state.
To obtain the spin triplet gap, we do similar things, but target
states with total $S_z = 0$ and $S_z = 1$ separately.  As for the
staggered magnetization, we perform a second order polynomial
extrapolation of the singlet and triplet gaps to the thermodynamic
limit (Figs.~\ref{Fig:StrFacSpinGap}(c,d)).  Consistent with
expectation, both $\Delta_S(L=\infty)$ and $\Delta_T(L=\infty)$
vanish in the two AFM phases.  They are both, however, non-zero and
large in the intervening region (see Fig.~\ref{Fig:PhaseDiagram}).
This rules out any state with broken SU(2) spin symmetry.

We notice that the singlet gap remains consistently below the
triplet gap throughout the intermediate phase.  This is an
indication of short-range singlet formation.   It is consistent with
a spin liquid state, and with a system with weak VBS order.  We
would, however, expect a strong VBS state to have a triplon
excitation, corresponding to breaking one singlet bond, as the
lowest energy bulk excitation, lower than singlet excitations which
require breaking two singlets.  So we can exclude a strong VBS state
in this sense based on the excitation spectrum.

\section{Topological Entanglement Entropy }
\label{sec:topol-entang-entr}

The above results provide evidence {\sl against} conventional
ordering in the intermediate region.  Magnetic ordering appears
comfortably excluded by both the correlation function and excitation
spectrum analysis.   Extrapolation of the dimer correlations to the
thermodynamic limit argues that VBS order is absent as well, but we
cannot exclude some very weak ordering on these grounds alone.

We now undertake a {\sl positive} evidence for a QSL with
topological order -- the topological entanglement entropy. The
topological entanglement entropy is obtained from the von Neumann
entanglement entropy $S(A)$.  The latter is defined for a state
$|\psi_0\rangle$ (which we take to be the ground state) and a
partition of the full system into a subsystem $A$ and its complement
$B$, by first constructing the reduced density matrix $\rho_A = {\rm
Tr}_B |\psi_0\rangle \langle \psi_0 |$.  Then the entanglement
entropy $S(A) = - {\rm Tr}_A (\rho_A \ln \rho_A)$.  For a system
with a gap to all bulk excitations (as we have verified in
Sec.~\ref{sec:energy-gaps}), provided the boundary between $A$ and
$B$ is taken to be smooth (i.e. have no corners), the entanglement
entropy must scale according to
\begin{equation}
  \label{eq:34}
  S(A) \sim \sigma L - \gamma+ \cdots,
\end{equation}
where the omitted terms vanish in the large $L$ limit.  Here
$\sigma$ is a non-universal number that measures the local
entanglement across the boundary.  According to
Refs.~\onlinecite{Kitaev2006,Levin2006}, the {\sl positive} term
$\gamma$ is {\sl universal} constant reduction from the area
law.\cite{Kitaev2006,Levin2006} It arises entirely from non-local
entanglement, and is topological in origin.  In particular, the area
law is {\sl strictly} obeyed, i.e. $\gamma=0$, for any state without
long-range entanglement, that is, which can be smoothly deformed
into a product state.  This is true, in the absence of spontaneously
broken symmetry, for any ground state which does not exhibit
topological order, i.e. which is not a topological
QSL.\cite{Kitaev2006,Levin2006}\

Although it is not discussed in the seminal papers on topological
entanglement entropy, a non-zero {\sl negative} $\gamma$ (i.e. a
positive correction to the area law) {\sl can} arise from discrete
spontaneous symmetry breaking (more severe positive corrections to
the area law arise in the case of a continuous broken
symmetry,\cite{2011arXiv1112.5166M,PhysRevB.84.165134} but this is
inconsistent with the existence of a gap to bulk excitations).  In
particular, in an ideal model with an exact discrete symmetry of the
Hamiltonian, the eigenstates must form irreducible representations
of the symmetry group.  For simple abelian groups such as $Z_N$,
these representations are one dimensional, so this implies the
Hamiltonian eigenstates are mutual eigenstates of the symmetry
generators.  This applies of course to the absolute ground state of
the system, which is therefore a Schr\"odinger cat state, which
superimposes the symmetry broken global ground states with equal
weight.  For the case of a fully broken $Z_N$ symmetry, with $N$
degenerate ground states in the thermodynamic limit, this gives rise
to $N$ terms in the Schmidt decomposition of the ground state, and
therefore a correction $\gamma=-\ln(N)$, i.e. a positive correction
to the area law or $\ln(N)$.  We have indeed observed such behavior
numerically in test studies of the simplest quantum transverse field
Ising model in the ferromagnetic phase, consistent with the expected
$\gamma = -\ln(2)$ for this case.

Thus we see that there are two potential sources of a non-zero
constant term in the entanglement entropy.  A topological
contribution which {\sl
  decreases} the entropy, and a symmetry breaking contribution which
{\sl increases} it. The latter correction arises from {\sl global}
entanglement of the entire system.  In work completed since the
earlier version of this article
appeared,\cite{jiang12:_ident_topol_order_entan_entrop} it has been
shown that the DMRG, which is a minimum entanglement approximation,
tends to converge, for large systems, to quasi-ground states which
capture all entanglement out to a long length scale, but not the
last global entanglement.  That is, for long systems, the
convergence of the DMRG is first to a {\sl Minimum
  Entanglement State} (MES) amongst the manifold of states comprising
the degenerate ground states in the thermodynamic
limit\cite{jiang12:_ident_topol_order_entan_entrop}.  For
topologically ordered phases, which have a ground state degeneracy
in the thermodynamic limit of topological origin, the MES exhibits
the universal reduction of entanglement entropy, i.e. the universal
positive value of $\gamma$.  For symmetry broken states, for which
there is a ground state degeneracy in the thermodynamic limit
dictated by symmetry, the MES is simply a single product-like state,
with $\gamma=0$. As shown by Jiang, Wang and Balents in
Ref.~\onlinecite{jiang12:_ident_topol_order_entan_entrop}, for a
fixed system size which is not too large, the DMRG can be pushed to
converge to the global ground state, by increasing the number of
states $m$. This is accompanied in these cases by a sharp {\sl
increase} in the entanglement entropy.  By increasing the length
$L_x$ of the system at fixed $L_y$, this final increase in the
entanglement entropy can be pushed beyond the range of feasible
calculations, and the simulation is guaranteed to obtain the MES. In
the MES, the constant correction $\gamma$ is entirely of topological
origin, and is zero in discrete symmetry breaking states.  Thus in
this limit $\gamma$ {\sl is} the topological entanglement entropy,
and a non-zero result proves that the state is a (topological) QSL.
Moreover, we see from the above discussion that a positive $\gamma$
can only come from topological order, so we do not obtain false
positive signatures of topological order from symmetry breaking.

\begin{figure}
    \includegraphics[width=3.2in]{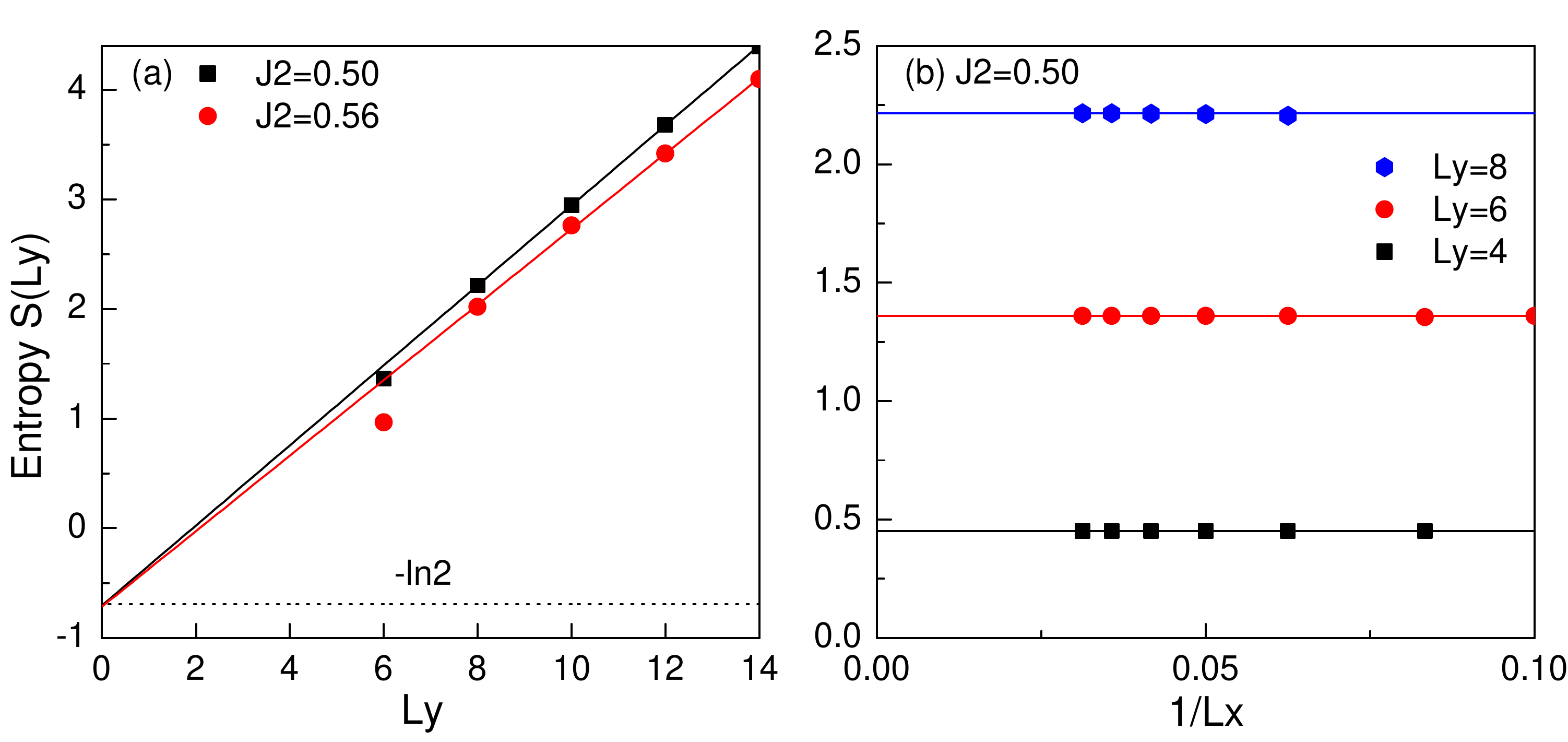}
    \caption{(Color online) The entanglement entropy at $J_2=0.5$ and
        $0.56$. (a) The entanglement entropy $S(L_y)$ for
      $L_y=6-14$. By fitting $S(L_y)=aL_y-\gamma$, we obtain $\gamma\sim
      0.70\pm 0.02$ for $J_2=0.5$, and $\gamma\sim 0.72\pm 0.04$ for
      $J_2=0.56$. (b) Length dependence of the entanglement entropy for
      $J_2=0.50$ and several system widths.  One observes that the
      entropy is almost independent of $L_x$ for long systems (a small
      increase with $L_x$ can be observed for the smallest $L_x$ at
      $L_y=8$).  } \label{Fig:Entropy}
\end{figure}

In Fig.\ref{Fig:Entropy}(a), we plot von Neumann entanglement
entropy $S(L_y)$ associated with the constant $x$ cut which
separates the cylinder into two symmetric parts of equal length,
$L_x/2$, as a function of $L_y$, with $L_y$ even (for $L_y$ odd,
there are additional effects which we discuss in
Sec.~\ref{sec:odd-even-effect}).  By comparing systems of different
lengths (Fig.~\ref{Fig:Entropy}b), we see that the entropy is
essentially independent of $L_x$ for $L_x>2L_y$, and so equal to its
limit at $L_x=\infty$. We then extrapolate $\gamma$ from the fitting
function $S(L_y)=a L_y-\gamma$. For $J_2=0.5$, deep in the
magnetically disordered phase, our results show that $\gamma=
0.70\pm 0.02$.  This value appears constant, within numerical
uncertainty, within the intermediate phase: for $J_2=0.56$ (close to
the quantum phase transition point $J_2=0.62$), we obtain in the
same way $\gamma =0.72\pm 0.04$.  Without even consider the
magnitude of $\gamma$, the fact that we see a negative rather than
positive correction to the entanglement entropy is strong evidence
against VBS order.

As mentioned in the Introduction, the topological entanglement
entropy $\gamma$ takes discrete values in topological phases.  The
minimum possible value for systems with unbroken time-reversal
symmetry is $\gamma=\ln (2) \approx 0.69$, which is within 2\% of
the numerical results.  The constancy of the numerical topological
entanglement entropy and the consistency with the theoretically
allowed value of $\ln(2)$ constitute strong evidence for a
topological QSL state.  The appearance of the pure number $\ln(2)$
(within happily small numerical uncertainty, of course) is certainly
very striking coming out entirely unsolicited from the DMRG
calculations.

Notably, $\gamma = \ln(2)$ is the expected value for a $Z_2$ QSL
phase.  The $Z_2$ QSL is in many ways the simplest spin-liquid
state, and has appeared repeatedly in theories of quantum magnets.
As a rather complete theory of the low energy properties of  $Z_2$
QSLs is available, we can compare this to numerics in various ways.

\section{Odd-even effect}
\label{sec:odd-even-effect}

In this section, we make such a comparison based on the theory of
the $Z_2$ QSL.  Specifically, in a $Z_2$ QSL on the square lattice,
it is predicted that cylinders with odd circumference -- and not
those with even circumference -- should exhibit non-vanishing bulk
staggered dimerization.  This even-odd effect was first obtained, to
our knowledge in Ref.\onlinecite{note}, by analysis of quantum dimer
models\cite{Rokhsar1988,Moessner2001}. Specifically, for a long
cylinder with (even) $L_x\rightarrow \infty$ and odd $L_y$, the
$Z_2$ QSL induces a non-vanishing staggered dimerization,
\begin{equation}
  \label{eq:35}
\langle B_i^x\rangle=\overline{B^x}+ D_x (-1)^{x_i},
\end{equation}
with $D_x \sim e^{-L_y/\tilde\xi}$ exponentially decreasing with
circumference.  By contrast, no dimerization appears for even $L_y$.
We obtain this behavior in Appendix~\ref{sec:stagg-dimer} directly
from the effective $Z_2$ gauge theory description, which shows that
it is a {\sl
  universal} feature of $Z_2$ QSLs on the square lattice, and not
particular to the quantum dimer models studied in Ref.\cite{note}.

Precisely this behavior is observed in our numerics.
Fig.~\ref{Fig:DimerCheck}(a,b) contrast the oscillatory and
non-oscillatory horizontal bond expectation values obtained for odd
and even $L_y$.  For even $L_y$, some small boundary effects are
observed, decaying over $\sim 3$ lattice spacings.
Fig.\ref{Fig:DimerCheck}(c) shows the exponential behavior of $D_x$
obtained as the difference of even and odd bonds at the center of
the sample.  Interestingly, theories predict (see Ref.\onlinecite{note} and also Supplementary
Information) $\tilde\xi=2\xi$, where $\xi$ is the true dimer
correlation length defined through the dimer correlation function.
This explains the rather slow decay of $D_x$, which fits to
$\xi\approx 5$, reasonably consistent with $\xi_d \approx 4$ found
(see Sec.~\ref{sec:corr-funct}) from the examination of VBS
correlation functions.  While some even-odd effect might be expected
in a columnar dimer phase for narrow cylinders, the
exponentially-decaying behavior and results of other tests (see
Sec.~\ref{sec:boundary-effects}) seem consistent only with a $Z_2$
QSL.

\section{Discussion}
\label{sec:discussion}

The previous sections have shown that DMRG makes a compelling case
for a non-magnetic intermediate state in the $J_1$-$J_2$ model.
From direct measurements of the dimer order parameter and
correlations, the intermediate state appears to have no or very weak
VBS order.  Most dramatically, we find a robust constant suppression
of the entanglement entropy relative to the generic area law, known
as {\sl topological entanglement
  entropy}, which is a unequivocal
signature of topological order.    The value of the topological
entanglement entropy we find is within 2\% (and our numerical
uncertainty) of the expected universal value $\gamma=\ln(2)$ for the
simplest $Z_2$ QSL state, which suggests comparison of specific
theoretical prediction for this $Z_2$ phase to numerics.  We indeed
find a characteristic even-odd effect in the staggered dimerization,
consistent with this state.

It is worth noting that ours is not the only suggestion of a QSL
state in the $J_1$-$J_2$ model.  Notably, after the initial version
of this paper appeared, a parallel work\cite{Wang2011T} came to
similar conclusions based on a tensor network variational method.

\subsection{Could this be a weak VBS state with strong finite size effects?}
\label{sec:boundary-effects}

In our opinion the above results all point in the same direction,
and are especially definitive given the seemingly unassailable
implication of the observed topological entanglement entropy.
Nevertheless, following an earlier version of this paper,
Sandvik\cite{PhysRevB.85.134407} has suggested, by comparison with
quantum Monte Carlo results for so-called J-Q models on cylinders,
that similar behavior might occur for a system with a VBS ground
state in the thermodynamic limit, due to strong finite size effects.
We discuss this suggestion here.

\subsubsection{Difference of models}
\label{sec:difference-models}

The results of Ref.~\onlinecite{PhysRevB.85.134407} are based on the
J-Q models, which have four or six spin interactions (with
coefficient $Q$).  These multi-spin interactions {\sl explicitly}
involve interactions between dimers, and as a consequence rather
naturally favor VBS states.  For instance, the simplest mean-field
treatment of the Q term in the J-Q$_2$ model would proceed from by
decoupling it by defining a mean-field dimer expectation value of
the dimer operator, and thereby a VBS phase appears when the Q term
becomes substantial.  Thus it is natural and intuitive to expect a
VBS phase in the J-Q models.  By contrast, there is no a priori
reason to expect dimer order in the $J_1$-$J_2$ model.  The notion
that a VBS state is somehow the most ``likely'' candidate for the
intermediate non-magnetic state in the $J_1$-$J_2$ case is a
misleading starting point.  More importantly, we should be cautious
in drawing conclusions from the J-Q models on the behavior of the
$J_1$-$J_2$ model.

\subsubsection{Entanglement entropy}
\label{sec:entanglement-entropy}

The most direct evidence for a QSL state we have obtained is the
topological entanglement entropy, remarkably close to the universal
expected value for a $Z_2$ QSL.  In
Ref.~\onlinecite{PhysRevB.85.134407}, Sandvik suggests that ``it
would not be surprising'' if a system near a N\'eel to VBS
transition (i.e. a DQCP) would exhibit a constant correction to the
area law similar to that expected for a topological phase.  In fact,
we have shown theoretically that in a VBS state, there is indeed a
constant correction {\sl but of opposite sign} to that of a
topological phase.  Thus even forgetting the magnitude of $\gamma$,
the sign alone is a strong argument against VBS order.  The fact
that the measured $\gamma$ is within 2\% of the very beautiful and
universal expected result $\ln(2)$ makes it hard to imagine this is
mere coincidence.

In the context of a putative DQCP, the constant correction for a VBS
state obtains if the system size is larger than the ``deconfinement
length'', below which there is an emergent U(1) symmetry unifying
the plaquette and columnar VBS states, and linear combinations in
between, with one another.  Would one perhaps see a signal similar
to the topological entanglement entropy were this length longer than
the system size?  Actually in this case we expect that the system
should appear to exhibit a gapless Goldstone mode, characteristic of
spontaneously breaking this U(1) symmetry.  This is the situation
discussed in
Refs.\onlinecite{2011arXiv1112.5166M,PhysRevB.84.165134}.  In fact
the behavior of the entanglement entropy in this case is even
further from that of a topological phase: a {\sl positive}
logarithmic enhancement of the entanglement entropy beyond the area
law is predicted, again of opposite sign to the topological case.
Moreover, in the 1d limit, $L_x \gg L_y$, the system should behave
as a 1+1-dimensional conformal field theory with central charge
$c=1$, and hence exhibit a logarithmic growth of entanglement
entropy, $S(L) \sim \frac{1}{6} \ln (L_x)$, in such a case.  This is
completely at odds with our observations -- observe the constant
behavior versus $L_x$ in Fig.~\ref{Fig:Entropy}b.

\subsubsection{VBS scaling}
\label{sec:vbs-scaling}

In Fig.~23 of Ref.\onlinecite{PhysRevB.85.134407}, our data for the
dimerization is replotted along with data for the $J$-$Q_2$ model on
a log-log plot, to fit to a single pure power law.  Data for $g(=J_2/J_1)=0.5$
is compared to $D_y^2 \sim L^{-\alpha}$ with $\alpha \approx 1.8$,
and for $g=0.56$ is slightly above it.  Small details of the data
for the latter case for the smallest systems, $L_x=4,6$, are used to
conclude that the system is VBS ordered in the infinite-size limit.
We disagree.  First, note the simple fact that the dimerization for
the $J_1$-$J_2$ model is much smaller than that of the $J$-$Q_2$
model (which is the more weakly VBS ordered of the two J-Q models).
Second, the scaling on this plot for the $J_1$-$J_2$ model (unlike
the J-Q models) is quite close to $\alpha=2$, which is, as mentioned
in the same paragraph of Ref.\onlinecite{PhysRevB.85.134407},
exactly the behavior expected for a {\sl non}-VBS phase.

\subsubsection{Even-odd effects in VBS states}
\label{sec:even-odd-effects}

One of the pieces of evidence for the $Z_2$ QSL state taken from our
numerics was the very distinct behavior of the staggered
dimerization in even and odd circumference systems, described in
Sec.~\ref{sec:odd-even-effect}.  While this is certainly consistent
with a $Z_2$ state, one could imagine similar behavior arising in a
system with VBS order in the thermodynamic limit.  Here we consider
the expected behavior in such a situation more carefully, for
comparison to our results.

\begin{figure}
    \includegraphics[width=3.2in]{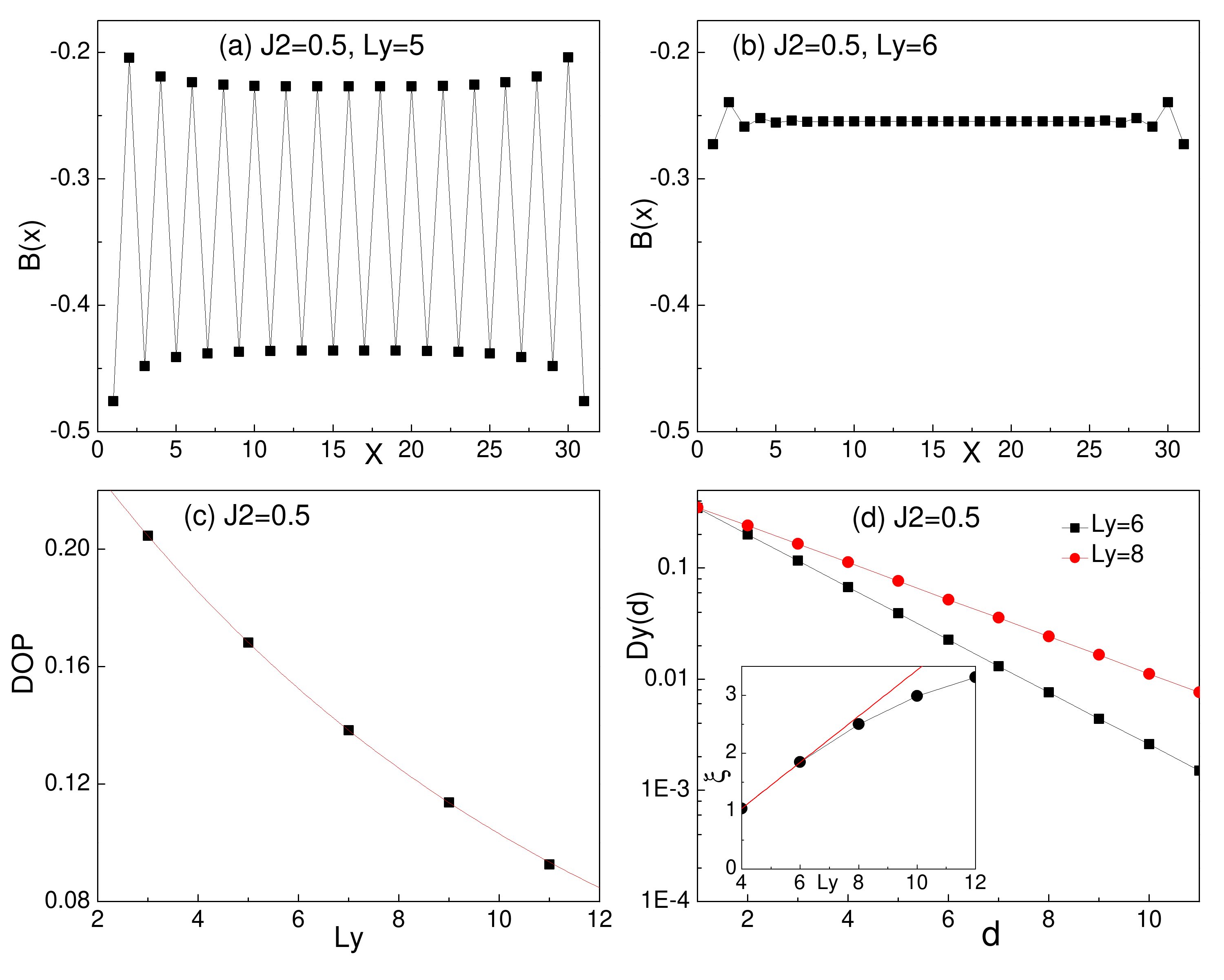}
\caption{(color online) Even-odd effect.  (a) Expectation value of
  horizontal bond operator, $\langle B_{i}^{x}\rangle$,  for $L_y=5$,
  $L_x=32$.  (b) The same expectation value for $L_y=6$, $L_x=32$. (c)
  Dimer order parameter $D_{d,\hat x}$ for odd $L_y$ at
$L_x=\infty$.  The red line denotes the exponential-decaying fitting
function with the form in Eq.(\ref{eq:27}). (d) Modified boundary
induced dimer order parameter for $Ly=6,8$, with $d$ the distance
from the boundary. Here the dimer order parameter is defined as the
dimer density difference between two nearest neighbor vertical dimer
bonds. Inset shows the correlation length $\xi$ along the cylinder
as a function of $L_y$.} \label{Fig:DimerCheck}
\end{figure}

Consider a system which is spontaneously dimerized in the 2d limit,
with a columnar dimer ground state.  This state is four-fold
degenerate, with four ground states consisting of two states with
``horizontal dimers'' staggered along the $x$ direction ($(\pi,0)$
order), and two states with ``vertical dimers'' staggered along the
$y$ direction ($(0,\pi)$ order).  In the thermodynamic limit, these
states are degenerate by rotation and translation symmetry.  When
confined to a cylinder, the anisotropy of the boundary conditions
breaks the symmetry between the horizontal and vertical states.  For
the case of odd-width cylinders, the vertical dimerization is
frustrated, because alternating ``rows'' of vertical dimers do not
fit into the sample.  This clearly would favor the horizontal dimer
states.  Amongst the two horizontal dimer states, the presence of an
end to the system splits the remaining degeneracy, so all degeneracy
is broken and we would expect long-range horizontal dimer order to
appear.  To this extent, the behavior for odd-width cylinders is the
same as observed in our numerics, and as expected for the $Z_2$ QSL.
The difference is in the scaling.  If the 2d system has a gapped
dimer ground state, we would expect the expectation value of the
dimerization to converge exponentially to a non-zero two dimensional
limit as the width of the cylinder increases, i.e.
\begin{equation}
  \label{eq:27}
  \left. D_x \right|_{\textrm{2d dimer state}} \sim \overline{D}_\infty + A e^{- L_y/\tilde\xi},
\end{equation}
where $A$ and $\tilde\xi$ are constants, and $\overline{D}_\infty$
is the value of the dimer order parameter in the thermodynamic
limit.

As shown in Fig.\ref{Fig:DimerCheck}(c), the numerical fitting to this
form gives $\overline{D}_\infty=0$ within numerical accuracy.  This is
entirely consistent with vanishing VBS order and a $Z_2$ QSL in the
thermodynamic limit, but of course cannot exclude some weak
dimerization smaller than our numerical uncertainties.  It seems to us
natural to take the former interpretation, since it is simpler.
According to the theory for the $Z_2$ QSL discussed in
Appendix~\ref{sec:stagg-dimer} and quantum dimer model
results\cite{note}, an exponential decay for the staggered
dimerization is expected with ``doubled'' correlation length
$\tilde\xi= 2\xi$, where $\xi$ is the true VBS correlation length.  We
find $\tilde\xi\approx10$ lattice spacings, which is equivalent via
Eq.~\eqref{eq:33} to $\xi\approx 5$.

For the even circumference cylinders, the vertical dimer order is
unfrustrated, and it is an energetic question, which likely depends
upon the details of the model, whether the vertical or horizontal
dimer order would be favored in this case.  If the horizontal dimer
state is favored, then we again expect behavior like
Eq.~\eqref{eq:27}, which is manifestly inconsistent with our numerics,
and markedly different from the $Z_2$ QSL.  However, it is perfectly
conceivable that the vertical dimer pattern is favored instead.  If
so, the periodic boundary conditions do not break the symmetry between
the two vertical dimer states, and so we expect the DMRG to converge
to the symmetric linear combination of the two dimer states, which
lacks any spontaneous dimer pattern. So at least the presence of an
even-odd effect in the static dimerization is consistent with a VBS
state, if the cylindrical geometry favors the two VBS states with
horizontal rows of vertical dimers.  On the face of it, this appears
consistent with our numerical results {\sl for the staggered
  dimerization}, if one assumes that the value of the dimerization
itself (extrapolated from odd circumference cylinders) is smaller than
our numerical uncertainty.  But it is worth pointing out that for this
scenario to hold, the even circumference system must be in a
Schr\"odinger cat state, and should exhibit a positive $\ln(2)$
enhancement of the entanglement entropy (negative TEE) as a
consequence, and moreover convergence to such a state should be
progressively more difficult with increasing $L_x$.  This is not at
all what we see.

\subsubsection{End effects}
\label{sec:end-effects}

In Ref.~\onlinecite{PhysRevB.85.134407}, strong boundary effects are
observed on the dimerization in the J-Q models.  Indeed, on symmetry
grounds, an open end breaks translation and reflection symmetries in
the $x$ direction, and as such should act as a ``boundary field'' on
the staggered dimer order $D_x$, i.e. it induces a term $- \lambda
D_x(x=0)$ in a Landau theory of this order.  On these grounds, we
always expect some staggered dimer order near the boundary.  If it
is energetically disfavored in the bulk, this will decay rapidly.
Otherwise, it will penetrate deep into the bulk.  In the J-Q models,
it was found that the boundaries induce a quite strong dimerization,
so that {\sl for even $L_y$} the bond expectation values $\langle
B_i^x\rangle$ oscillate visibly (c.f. in the inset of Fig.~6, and in
Fig.~15a of Ref.\onlinecite{PhysRevB.85.134407}, the bond
expectation value shows oscillations with large amplitude in the
J-Q$_3$ and J-Q$_2$ models, respectively).  By contrast, in the
$J_1$-$J_2$ model, we see in Fig.~\ref{Fig:DimerCheck}(b) that there
are no visible oscillations in the same quantity when $L_y$ is even.
This qualitative difference tells us that $D_x$ order is clearly
much less favorable in the $J_1$-$J_2$ model.

We next try to address the possibility, raised above, that the
cylindrical geometry when $L_y$ is even favors $D_y$ order, i.e.
horizontal rows of vertical dimers.  This is at odds with our
measurements of the dimer correlations and the entanglement entropy.
Still, it is more compelling to explicitly test to rule out the
possibility directly.  To do so, we have studied several modified
cylinders with even circumference, in which the ends of the cylinder
have been altered, breaking translational symmetry along $y$ in
order to break the degeneracy and favor one of the two vertical
dimer states. What we observe is that in all cases, as shown in
Fig.\ref{Fig:DimerCheck}(d), although dimer order is induced by this
symmetry breaking in the vicinity of the boundary, it decays
exponentially into the bulk of the cylinder.  The correlation length
$\xi_v$ for this vertical dimer order still depends on circumference
for the system sizes in our study, so we plot it versus $L_y$ to see
if it is limited by the system size (it does not appear to be), and
to extrapolate from this its value in the thermodynamic limit.  We
observe that this correlation length grows sub-linearly in $L_y$,
and extrapolates to $\xi_v\sim 4$ in the 2D limit (i.e.,
$L_y=\infty$). This is very different from what would be expected
for a 2d state with long-range dimer order, in which the non-zero
stiffness (surface tension) of the ordered dimer state would prevent
such decay (we would expect $\xi_v = \infty$ in this case).  If one
were to imagine that the system were proximate to a DQCP, and $L_y$
were smaller than the deconfinement length, then we would instead
expect $\xi_v \propto L_y$, which again is not consistent with our
results.  Note also that the value for $\xi_v$ is quite consistent
with the value for $\tilde\xi$ obtained earlier.  The fact that
vertical dimer order decays away, even when the most favorable
conditions have been created for it, is strong evidence against VBS
order in the 2d limit.

\subsection{Summary and Open Issues}
\label{sec:summary}

In conclusion, we have presented compelling evidence from accurate
DMRG calculations for a topological QSL state in the two dimensional
$J_1$-$J_2$ Heisenberg model.  This is the simplest example of such
a QSL discovered to date, and the only one to our knowledge for a
Heisenberg model on a Bravais lattice.  As such, it is particularly
attractive for further theoretical and experimental study.  We
anticipate, for instance, that our discovery will afford an
opportunity to explore the QSL mechanism of unconventional
superconductivity\cite{Anderson1987,Kivelson1987} in a controlled
theoretical setting.

Another consequence of topological order is the presence of
quasi-degenerate ground states on the torus or cylinder.  A two-fold
quasi-degeneracy is expected for a $Z_2$ QSL on the cylinder studied
here, with a splitting of order $L_x e^{-L_y/\xi}$ in the case of long
cylinders, where $\xi$ is the spin-spin correlation length (see
Appendix~\ref{sec:ground-state-degen}).  As shown in
Ref.\onlinecite{jiang12:_ident_topol_order_entan_entrop} and discussed
in Sec.~\ref{sec:topol-entang-entr}, the DMRG
preferentially converges, however, to just {\sl one} of the quasi-degenerate
ground states (specifically, a minimally entangled state).  This
explains the absence of an observed topological degeneracy in this and
other DMRG studies.\cite{White2011Kagome,
  Jiang2008Kagome,jiang12:_ident_topol_order_entan_entrop} It is a
non-trivial and open problem to obtain the second ground state and
thereby extract the topological energy splitting.  It is our
expectation that it is actually orders of magnitude smaller than the
bulk energy gaps.

The nature of the quantum phase transitions from the QSL to N\'eel
and striped antiferromagnetic phases is an interesting topic for
future study.  Though we have not focused on the transitions
themselves, and more work is clearly required to make strong
conclusions about them numerically, it appears that the transition
from the N\'eel to QSL state may be continuous.
Ref.\onlinecite{PhysRevB.85.134407}\ erroneously claims that a
N\'eel to QSL transition might be in the same universality class as
the DQCP between N\'eel and VBS order, because ``the operator
causing the VBS order is dangerously invariant''.  Though at the
DQCP the operator which {\sl
  distinguishes} between columnar and plaquette VBS order is
dangerously {\sl irrelevant}, even when this operator's coefficient
in the Hamiltonian is tuned to zero, the non-magnetic phase has
spontaneous VBS order.  So this claim is incorrect.  In fact, such a
transition requires an entirely different theory. A novel suggestion
for the theory of this critical point has been made in
Ref.\onlinecite{2012arXiv1204.5486M}, and it would be interesting to
compare it to further numerical studies.

\begin{acknowledgements} We would like to thank Cenke Xu, Ashvin Vishwanath, Zheng-Yu Weng,
 Steve White, Zhenyue Zhu, Max A. Metlitski, Dong-Ning
Sheng, Zheng-Cheng Gu and H.Y. especially thanks Steve Kivelson for
inspiring discussions. This work was supported in part by the NBRPC
(973 Program) 2011CBA00300 (2011CBA00302). H.C.J. sincerely thanks
the hospitality of Microsoft Station Q, where part of the numerical
simulation was done on the Cirrus cluster. H.Y. was supported by NSF
Grant DMR-0904264 at Stanford. L.B. and H.C.J. were supported by NSF
grant DMR-0804564. This work was partially supported by the the KITP
NSF grant PHY05-51164 and the NSF MRSEC Program under Award No. DMR
1121053.
\end{acknowledgements}


\appendix

\section{$Z_2$ gauge theory}
\label{sec:symmetries}

Here we discuss an effective $Z_2$ gauge theory description of the
QSL state,\cite{senthil2000z} and in particular derive the behavior
of the dimerization and ground state quasi-degeneracy discussed in
the main text.  We begin with the Hamiltonian
\begin{eqnarray}
  \label{eq:2}
  H  & = &   -K \sum_{\Box}  \prod_{\langle ij\rangle \in \Box}
  \sigma_{ij}^z - h \sum_{\langle ij\rangle} \sigma_{ij}^x + r
  \sum_i n_i \\ & &
  -
  \sum_{\langle ij\rangle} \sigma_{ij}^z \Big[ t \, b_{i\alpha}^\dagger
  b_{j\alpha}^{\vphantom\dagger} + \Delta \eta_i \left(b_{i\alpha}
  \epsilon_{\alpha\beta} b_{j\alpha} + {\rm h.c.}\right)\Big], \nonumber
\end{eqnarray}
where $n_i = b_{i\alpha}^\dagger b_{i\alpha}^{\vphantom\dagger}$ and
$\eta_i = (-1)^{x_i+y_i}$.  We introduced ``spinon'' operators
$b_{i\alpha}$ which transform as spinors under SU(2), and obey
standard commutation relations
$[b_{i\alpha}^{\vphantom\dagger},b_{j\beta}^\dagger]=
\delta_{ij}\delta_{\alpha\beta}$.  The physical spin operators are
related to them by ${\bf S}_i = \frac{1}{2}b_{i\alpha}^\dagger
{\boldmath{\sigma}}_{\alpha\beta}b_{i\beta}^{\vphantom\dagger}$. The
$\sigma_{ij}^z$ operators are Pauli matrix $Z_2$ gauge fields, which
we will refer to as the ``magnetic'' gauge fields.  $Z_2$ gauge
symmetry is enforced by the constraint
\begin{equation}
  \label{eq:3}
  \prod_{|j-i|=1} \sigma_{ij}^x =- (-1)^{n_i}.
\end{equation}
Note that the product in Eq.~(\ref{eq:3}) is over $j$ not $i$.  This
is the analog of Gauss' law for the ``electric'' field
$\sigma_{ij}^x$.  This constraint ``generates'' the Ising gauge
symmetry $\sigma_{ij}^z \rightarrow s_i s_j \sigma_{ij}^z$, $b_i
\rightarrow s_i b_i$, where $s_i = \pm 1$ can be chosen arbitrarily
for each site.

\subsection{Staggered dimerization}
\label{sec:stagg-dimer}

Here we obtain the behavior of the staggered dimerization from the
$Z_2$ gauge theory.  For this purpose, it is sufficient to integrate
out the spinons, since we discuss local properties of the QSL state
which has a spin gap (but see below
Sec.~\ref{sec:ground-state-degen}).   We can obtain this limit from
Eq.~(\ref{eq:2}) by taking $r$ large, which projects the problem
onto the subspace with $n_i=0$.  Then the Hamiltonian reduces to
\begin{equation}
  \label{eq:4}
  H  =   -K \sum_{\Box}  \prod_{\langle ij\rangle \in \Box}
  \sigma_{ij}^z - h \sum_{\langle ij\rangle} \sigma_{ij}^x ,
\end{equation}
and
\begin{equation}
  \label{eq:5}
  \prod_{|j-i|=1} \sigma_{ij}^x =-1.
\end{equation}
Eqs.~(\ref{eq:4},\ref{eq:5}) describe the ``odd Ising gauge
theory''. It is in the deconfined (QSL) phase for $K/h > x_c$, where
$x_c$ is some order one number specifying the critical point.

Now consider the staggered dimerization, $D_x = (-1)^{x_i}\langle
D_i^x\rangle - \langle D_{i+\hat{x}}^x\rangle$, defined in the main
text.  On symmetry grounds, we expect that $\langle D_i^x\rangle
\propto \langle \sigma_{i,i+\hat{x}}^x\rangle$ (this relation can
also be derived by perturbation theory in $t/r$).  We will derive
the odd/even effect for the staggered dimerization in finite-width
cylinders in two ways. First, we obtain it directly from the Ising
gauge theory in the strong coupling limit, which is a very short
derivation.  Second, we obtain it using duality and field theory,
which exposes the universal nature of the staggered dimerization and
its relation to $Z_2$ vortex (``vison'') excitations.

To see how one might expect the dimerization, we first consider the
``topological'' operator
\begin{equation}
  \label{eq:7}
  Q_x = \prod_{y=1}^{L_y} \sigma_{xy;x+1y}^x.
\end{equation}
This operator commutes with $H$ and is thus a constant of the
motion. Moreover, if we consider the case $x=1$ at the left hand
side of the system, we obtain
\begin{equation}
  \label{eq:8}
  Q_1 = \prod_{y=1}^{L_y} \left( \prod_{|j-i|=1} \sigma_{ij}^x
  \right)_{i=(1,y)} = (-1)^{L_y},
\end{equation}
where we have used Eq.~(\ref{eq:5}).  Again using Eq.~(\ref{eq:5}),
one obtains
\begin{equation}
  \label{eq:9}
  Q_x = (-1)^{x L_y}.
\end{equation}
Thus $Q_x=1$ for even $L_y$, but oscillates, $Q_x=(-1)^x$, for odd
$L_y$.  Although this is not the dimerization itself, it suggests
the presence of staggered dimerization in the case of odd $L_y$.

\subsubsection{Direct derivation}
\label{sec:direct-derivation}

We now turn to the first derivation, working deep in the deconfined
phase, taking $K\gg h$, and proceed by direct calculation
perturbatively in $h$.  For $h=0$, the ground state(s) are obtained
by simply choosing a classical configuration of $\sigma_{ij}^z$ with
zero Ising gauge flux, $\prod_{\langle ij\rangle \in \Box}
\sigma_{ij}^z=1$ on all plaquettes (for instance the state with
$\sigma_{ij}^z=+1$ on all bonds), and then projecting this state to
satisfy Eq.~(\ref{eq:5}):
\begin{equation}
  \label{eq:6}
  |\psi_0\rangle = \prod_i \hat{P}_i |\sigma_{ij}^z=1\rangle,
\end{equation}
where
\begin{equation}
  \label{eq:10}
  \hat{P}_i = \frac{1}{2} -
    \frac{1}{2}\prod_{|j-i|=1} \sigma_{ij}^x
\end{equation}
In this state, the expectation value of $\sigma_{ij}^x$ vanishes.
This can be seen as follows.  Define the Wilson loop operator
\begin{equation}
  \label{eq:11}
  W[{\mathcal C}]= \prod_{\langle ij\rangle \in \mathcal{C}} \sigma_{ij}^z,
\end{equation}
where $\mathcal{C}$ is a closed curve on the lattice.  All such
Wilson loops commute with the projectors $\hat{P}_i$, so
$|\psi_0\rangle$ is an eigenstate of the Wilson loop with
$W[\mathcal{C}]|\psi_0\rangle = |\psi_0\rangle$.  Moreover, since
$W[\mathcal{C}]^2=1$, we have
\begin{equation}
  \label{eq:12}
   \langle \psi_0|\sigma^x_{ij}|\psi_0\rangle = \langle
   \psi_0|W[\mathcal{C}]\, \sigma^x_{ij}\, W[\mathcal{C}]|\psi_0\rangle = -
   \langle \psi_0|\sigma^x_{ij}|\psi_0\rangle  = 0,
\end{equation}
if we choose $\mathcal{C}$ to be a curve containing the both
$\langle ij\rangle$.  To achieve a non-zero result, we must consider
non-zero orders of perturbation theory in $h/K$.  In general, the
form of the perturbative eigenstate is
\begin{equation}
  \label{eq:13}
  |\psi\rangle \propto \sum_{n=0}^\infty c_n
  \left[ \hat{R} H' \right]^n |\psi_0\rangle,
\end{equation}
where $\hat{R} = {\sf P} (E_0-H_0)^{-1} {\sf P}$ is the resolvent
with $H_0 = H (h=0)$ and $E_0$ the ground state energy of $H_0$) and
${\sf P}=1-|\psi_0\rangle\langle\psi_0|$ is the projector onto the
unperturbed excited state subspace, $H'=H-H_0= - h \sum_{\langle
ij\rangle} \sigma_{ij}^x$, and the $c_n$ are numerical coefficients.
This can be expanded to give a series of terms, each involving a
product of $n$ electric gauge fields acting on $|\psi_0\rangle$ at
$O[(h/K)^n]$.  For each such term, we can repeat the argument in
Eq.~(\ref{eq:12}).  We will achieve a vanishing result provided we
can choose $\mathcal{C}$ to contain an odd number of links that
coincide with the set of links $\mathcal{L}$ containing the electric
fields in the corresponding term in the wavefunction {\sl
  and} the link $\langle ij\rangle$ in the expectation value.  This is
always possible unless the ``dual'' of $\mathcal{L}$ forms a closed
loop.  This dual is formed by associating a link of the dual lattice
with each link in $\mathcal{L}$.  If the dual of $\mathcal{L}$
indeed forms a closed loop, then the closed loop $\mathcal{C}$ must
intersect it an even number of times.

Thus we obtain non-zero contributions only from terms in which
$\mathcal{L}$ is comprised of closed dual loops.  There are trivial
contributions from short loops, the minimal one being the case when
$\mathcal{L}$ contains $\langle ij\rangle$ twice, which is first
order in $h/K$.  This gives a non-zero constant contribution to the
expectation value, but one which is {\sl uniform}, and hence does
not
correspond to a staggered dimerization.  
A non-trivial result is obtained first at $O[(h/K)^{L_y-1}]$, from
the smallest closed dual loop encircling the cylinder and containing
the bond due to $\langle ij\rangle$, which must be a horizontal
bond.  This leading term arises from the $O[(h/K)^{m}]$ correction
to the ground state ket and the $O[(h/K)^{L_y-1-m}]$ correction to
the ground state bra ($m=0,1,\cdots,L_y-1$), giving
\begin{eqnarray}
  \label{eq:14}
&&   \langle \sigma_{ii+\hat{x}}^x \rangle = \\
&& \cdots + C_n
  \sum_{m=0}^{L_y-1} {L_y-1\choose m}\left(\frac{h}{K}\right)^{(L_y-1)} \langle
  \psi_0|Q_{x_i}|\psi_0\rangle
  \nonumber\\
  & &= \cdots + C_n
  \left(\frac{2h}{K}\right)^{L_y-1} (-1)^{L_y x}. \nonumber
\end{eqnarray}
Here $C_n$ is a numerical coefficient which should be determined
from a more refined analysis.  We therefore conclude that for odd
$L_y$, we obtain the staggered dimerization discussed in the main
text, with amplitude $D_x \sim (2h/K)^{L_y-1} = \exp[- \ln(K/2h)
(L_y-1)]$, exponentially decaying with circumference as advertised.
This result derived from the odd Ising gauge theory is qualitatively
consistent with the one obtained from the analysis\cite{note} of
quantum dimer models.

\subsubsection{Dual derivation}
\label{sec:dual-derivation}

While the above derivation is simple and direct, it relies on the
strong coupling expansion, which, although it is expected to be
qualitatively correct in the deconfined phase, is not obviously
general.  It is instructive to obtain the staggered dimerization by
a more circuitous dual route, which exposes the universality of the
result and gives a more direct physical picture.

The duality transformation of Eqs.~(\ref{eq:4},\ref{eq:5}) is
accomplished by defining
\begin{eqnarray}
  \label{eq:15}
  \tau_a^x & = & \prod_{\langle ij\rangle \in a} \sigma_{ij}^z, \\
  \label{eq:16} \sigma_{ij}^x & = & \mu_{ab}\tau_a^z \tau_b^z ,
\end{eqnarray}
where ${\boldmath \tau}_a$ are new Pauli matrices.  In
Eq.~(\ref{eq:15}) $\langle ij\rangle$ are the bonds associated with
dual site $a$ at the center of a direct plaquette, and in
Eq.~(\ref{eq:16}), the dual sites $a,b$ are those at the centers of
the two plaquettes neighboring the bond $\langle ij\rangle$.  The
scalars $\mu_{ab}$ must be chosed to satisfy Eq.~(\ref{eq:5}), which
requires that their product around a dual plaquette must equal $-1$.
The dual Hamiltonian is then a fully frustrated transverse field
Ising model:
\begin{equation}
  \label{eq:17}
  H = - h \sum_{\langle ab\rangle} \mu_{ab} \,\tau_a^z \tau_b^z - K
  \sum_a \tau_a^x.
\end{equation}
The $\tau_a^z$ operator has the physical interpretation of creating
an Ising vortex (vison) on plaquette $a$.  In the deconfined phase,
when $K/h > x_c$, the visons are gapped excitations in the
``paramagnetic'' phase of this dual Ising model.  We will see that
the dimerization is related to virtual vison excitations.

To see this, we obtain a continuum limit of Eq.~(\ref{eq:17}), valid
in the deconfined phase, as follows (qualitatively identical results
can be obtained in many other ways, for instance by an expansion
about mean field theory, or by strong coupling expansions).  It is
convenient to work in a path integral formulation in the $\tau_a^z$
basis, and ``soften'' the spins $\tau_a^z \rightarrow \varphi_a$.
The Euclidean action in the time continuum limit is then
\begin{eqnarray}
  \label{eq:18}
  S = \int \! d\tau \, \Big\{  && - h \sum_{\langle ab\rangle} \mu_{ab}\,
    \varphi_a \varphi_b  \\
    && + \sum_a \left[ \frac{\kappa}{2}
      (\partial_\tau \varphi_a)^2 + \frac{r}{2} \varphi_a^2 + u
      \varphi_a^4 \right]\Big\}, \nonumber
\end{eqnarray}
where $\kappa, r$ and $u$ are phenomenological parameters.  In the
deconfined phase, the fluctuations of $\varphi_a$ are small, and it
is sufficient to truncate the action to quadratic order.  The
dominant fluctuations are those near the minimum of the quadratic
form.  To find them, we must choose a gauge for the frustrated dual
exchange. It is convenient to make the following choice:
\begin{equation}
  \label{eq:19}
  \mu_{a,a+\hat{y}} = (-1)^{x_a}, \qquad \mu_{a,a+\hat{x}}=1.
\end{equation}
Here we have taken the dual lattice sites to have integer
coordinates. The unit cell in this gauge contains two sites.
Therefore, Fourier transforming to go to the Bloch basis, we obtain
the inverse Green's function describing the virtual fluctuations of
the visons,
\begin{equation}
  \label{eq:20}
  G^{-1} = (\kappa\,\omega_n^2 + r) {\sf I} -4h
  \begin{pmatrix}
    \cos k_y  & \cos k_x \\
    \cos k_x & -\cos k_y
  \end{pmatrix}.
\end{equation}
Here the ``magnetic'' Brillouin zone is $|k_x|\leq \pi/2, |k_y|\leq
\pi$.  The dominant fluctuations, corresponding to the minimum
eigenvalue of $G^{-1}$ ($= r - 4\sqrt{2}h$), occur at the two
inequivalent values $(k_x,k_y)=(0,0)$ and $(k_x,k_y)=(0,\pi)$.  The
corresponding eigenvectors are $\phi^{(1)}=(\cos \frac{\pi}{8},\sin
\frac{\pi}{8})$ at $k=(0,0)$ and $\phi^{(2)} =(\sin
\frac{\pi}{8},\cos \frac{\pi}{8})$ at $k=(0,\pi)$.  Focusing on
these lowest energy excitations, we therefore write
\begin{eqnarray}
  \label{eq:21}
  \varphi_a & \sim & \phi^{(1)}_a \Phi_1(x_a,y_a) + \phi^{(2)}_a
  (-1)^{y_a} \Phi_2(x_a,y_a),
\end{eqnarray}
where $\phi^{(i)}_a$ takes the two values of eigenvector $i$ given
above when $a$ is on the two distinct sublattices, and $\Phi_i(x,y)$
is a slowly-varying continuum field.  The bulk effective action is
then
\begin{eqnarray}
  \label{eq:22}
  S = \frac{\kappa}{2}\sum_{i=1,2}\int \! d\tau dx dy\, \left\{
    (\partial_\tau \Phi_i)^2 + v^2 (\nabla \Phi_i)^2 + m^2 \Phi_i^2\right\}.
\end{eqnarray}
This action describes two degenerate minimum energy vison states. It
was discussed first to our knowledge in Ref.\cite{PhysRevB.30.1362},
in the context of frustrated Ising models.  It is instructive to
express the VBS order parameter in terms of $\Phi_i$.  If we
consider the horizontal bonds,
\begin{eqnarray}
  \label{eq:1}
  D_x & = & (-1)^{x_i} \left( {\bf S}_i \cdot {\bf
    S}_{i+\hat{x}} -  {\bf S}_{i+\hat{x}} \cdot {\bf
    S}_{i+2\hat{x}}\right) \\
  & \sim & (-1)^{x_i}\left(  \sigma^x_{i,i+\hat{x}} -
  \sigma^x_{i+\hat{x},i+2\hat{x}}\right) \nonumber \\
  & \sim & (-1)^{x_a}\left( \tau_{a}^z \tau_{a+\hat{y}}^z +
  \tau_{a+\hat{x}}^z \tau_{a+\hat{x}+\hat{y}}^z\right) \nonumber\\
  & \sim & (c \Phi_1 + s\Phi_2)(c \Phi_1 - s \Phi_2) - (s\Phi_1 +
  c\Phi_2)(s\Phi_1 - c\Phi_2) \nonumber \\
  & \sim & \Phi_1^2-\Phi_2^2,
\nonumber
\end{eqnarray}
where in the penultimate line of Eq.~\eqref{eq:1}, $c=\cos \pi/8$
and $s=\sin \pi/8$.  By a similar calculation, one finds that the
vertical bond dimerization is
\begin{equation}
  \label{eq:23}
  D_y =  (-1)^{y_i} \left( {\bf S}_i \cdot {\bf
    S}_{i+\hat{y}} -  {\bf S}_{i+\hat{y}} \cdot {\bf
    S}_{i+2\hat{y}}\right) \sim 2 \Phi_1 \Phi_2.
\end{equation}
From this we obtain the result
\begin{equation}
  \label{eq:24}
  \Psi={D}_x+i {D}_y \sim (\Phi_1+ i \Phi_2)^2.
\end{equation}
The gauge invariant combination on the right hand side can thus be
identified as the familiar complex VBS order parameter $\Psi$.  This
result, and the action Eq.~\eqref{eq:22}, have been obtained many
times for quantum spin-$1/2$ systems on the square lattice.  Indeed,
both are largely independent of the microscopic model, and give the
{\sl minimal} set of excitations and their properties gives only the
assumptions of $Z_2$ topological order in the ground state and
half-integer spin per unit cell.  It would be interesting to
understand if other dimer patterns could in principle arise, if the
low energy vison states were selected from a different projective
symmetry group.\cite{PhysRevB.84.014402}  In two dimensions, in the
$Z_2$ QSL phase, there is no VBS order, so the visons are gapped and
the VBS order parameter $\Psi$ also is uncondensed, correspondingly.

We now consider the finite-size effects.  Taking periodic boundary
conditions on $\varphi_a$ in the $y$ direction imposes, using
Eq.~(\ref{eq:21}), periodic boundary conditions on $\Phi_1$ but {\sl
  anti-periodic} boundary conditions on $\Phi_2$ when $L_y$ is odd.  The
latter result can be readily understood in terms of the VBS order
parameter: on an odd-leg cylinder, the vertical component $D_y$ is
frustrated (staggering of rows of dimers does not ``fit'') and
should be antiperiodic, which requires $\Psi \rightarrow \Psi^*$
under the circuit around the cylinder, consistent with the
anti-periodic boundary conditions on $\Phi_2$.  Since the visons are
gapped, the antiperiodic boundary condition gives an exponentially
small effect in the thermodynamic limit, but it is non-zero and can
be readily calculated.

Regardless of boundary conditions, because $\Phi_1$ and $\Phi_2$ are
decoupled in Eq.~\eqref{eq:22}, $\langle D_y\rangle =0$, so there is
no VBS order of the vertical bonds.  The horizontal component,
however, is non-zero when $L_y$ is odd, so that the fields $\Phi_1$
and $\Phi_2$ are slightly inequivalent due to the boundary
conditions:
\begin{eqnarray}
  \label{eq:25}
  \langle D_x\rangle & \sim & \langle \Phi_1^2\rangle - \langle
  \Phi_2^2\rangle \\
  & \sim & \kappa^{-1} \int \! \frac{d\omega_n}{2\pi}
  \frac{dk_x}{2\pi} \Big[ \frac{1}{L_y} \sum_{k_y}  \,
  \frac{1}{\omega_n^2 + v^2 k^2 + m^2} \nonumber \\
  && -  \frac{1}{L_y} \sum^\prime_{k_y}  \,
  \frac{1}{\omega_n^2 + v^2 k^2 + m^2}\Big],\nonumber
\end{eqnarray}
where the first sum is over ``periodic'' momenta $k_y=2\pi n/L_y$,
and the second sum (with the prime) is over ``antiperiodic'' momenta
$k_y=2\pi (n+1/2)/L_y$, with integer $n$.  To proceed, we first
perform the frequency integration and then use the Poisson
resummation formula to obtain
\begin{eqnarray}
  \label{eq:26}
  \langle D_x\rangle & \sim & \frac{2}{\kappa} \sum_{p=0}^\infty \int \!
  \frac{dk_x}{2\pi} \int \frac{d k_y}{2\pi} \frac{\cos [(2p+1) k_y
    L_y]}{\sqrt{v^2 k^2 + m^2}} \\
  & \sim & \frac{2}{\pi\kappa v} \sum_{p=0}^\infty \int \!
  \frac{dk_x}{2\pi} M(k_x) K_0[(2p+1)M(k_x) L_y/v],\nonumber
\end{eqnarray}
where we carried out the $k_y$ integration in the last line, and
defined $M(k_x) = \sqrt{m^2 + v^2 k_x^2}$.  For large $L_y$, the
asymptotic form of the Bessel function can be used, $K_0(z) \sim
\sqrt{\pi/2z}e^{-z}$, and the dimerization is dominated by the $p=0$
term and the region $vk_x \ll m$ :
\begin{eqnarray}
  \label{eq:28}
  \langle D_x\rangle & \sim & \frac{2}{\pi\kappa v} \sqrt{\frac{\pi v}{m
      L_y}} \int \!  \frac{dk_x}{2\pi} \, m e^{-m L_y/v} e^{- v
    k_x^2 L_y/2m} \nonumber \\
  & \sim & \frac{\sqrt{2}m}{\pi \kappa v L_y} e^{- m L_y/v}.
\end{eqnarray}
As promised, we obtain exponential decay of the dimerization, and in
this case a prediction for the prefactor.  The physics of this
derivation is transparent: virtual fluctuations of $Z_2$ vortices
which propagate about the cylinder lead directly to the
dimerization. In this way we immediately see that this effect is
universal for $Z_2$ QSLs on the square lattice with $S=1/2$ spins.

Let us conclude this subsection with one remark on the dimer
correlation lengths.  The static dimerization on cylinders with odd
circumference decays with an apparent correlation length $\tilde\xi
= v/m$.  This is {\sl not} the same length which appears in the
dimer-dimer correlation function.  The latter is obtained from
correlation functions of $\Psi$, given in Eq.~\eqref{eq:24}.
Because the dimer order parameter $\Psi$ is quadratic in the vison
fields $\Phi_i$, and the $\Phi_i$ are Gaussian distributed, by
Wick's theorem the dimer-dimer correlation functions are {\sl
squares} of vison Green's functions.  Consequently, the exponential
decay of the dimer-dimer correlation function, which defines the
standard dimer correlation length $\xi$, is twice as fast, i.e.
\begin{equation}
  \label{eq:33}
  \tilde\xi=2\xi.
\end{equation}
This behavior has indeed been observed in the numerical studies in
the main text.

\subsection{Ground state degeneracy}
\label{sec:ground-state-degen}

It is well-known that the $Z_2$ spin liquid has degenerate ground
states in the thermodynamic limit on a cylinder or torus.  For the
cylindrical geometry studied here, two states are expected.  Here we
would like to understand the scaling of the gap between these two
states, and also better understand their character.  We will see
that, as discussed e.g. in Ref.\cite{PhysRevB.63.134521}, that the
presence of gapped spin excitations (which carry non-zero electric
gauge charge) makes a {\sl
  qualitative} difference in these properties.  This means that models
neglecting these excitations, in particular the very popular quantum
dimer models, actually give {\sl incorrect} or {\sl non-generic}
scaling for the finite-size quasi-degenerate gap.

Consider first the pure gauge theory, Eq.~\eqref{eq:4}, in which
coupling to matter fields is neglected.  The ground state degree of
freedom may be regarded as the presence or absence of a vison
through the hole in the cylinder.  The presence of the vison itself
is measured by the Wilson loop operator around the cylinder,
\begin{equation}
  \label{eq:29}
  W = \prod_{y=1}^L \sigma_{xy;xy+1}^z.
\end{equation}
A state with a $Z_2$ vortex in it has $W=-1$ and without has $W=1$.
However, the ground state will not be an eigenstate of $W$.  In
fact, consider the conjugate operator
\begin{equation}
  \label{eq:30}
  Q = \prod_{x=1}^L \sigma_{xy;xy+1}^x.
\end{equation}
This operator {\sl commutes} with $H$ defined in Eq.~\eqref{eq:4},
and so is a constant of the motion.  The two degenerate ground
states have $Q=\pm 1$ (we can pick any $y$, since others are related
by Eq.~\eqref{eq:5}).  Note that $WQ=-QW$, so an eigenstate of $Q$
is a symmetric or antisymmetric combination of the $W$ (vison)
eigenstates.  This indicates physically that the vison may tunnel
through the cylinder, by moving (virtually) through the entire long
length $L_y$ from one end to another, thereby connecting the $W=1$
and $W=-1$ states.  The tunneling amplitude for this process is
naturally expected to be exponential in the length of the event, so
we postulate that the gap in this case is $t_v \sim e^{-L_x/\xi_x}$.
This has been shown explicitly in many places in the literature.

This result is generic for the pure $Z_2$ gauge theory, and
continues to hold even if longer (but finite) range plaquette and
electric field terms are included.  It relies only on the fact that
$Q$ does not create any physical gauge flux through finite
plaquettes.  However, if a matter field (i.e. the spinons) is
present, the result is modified. To see this, let us imagine more
carefully integrating out the spinons in going from Eq.~\eqref{eq:2}
to Eq.~\eqref{eq:4}, for the case of a cylinder of finite
circumference.  Then we will obtain not only contributions from
small loops (which renormalize $K$ etc.), but also, occuring first
at $O(t^{L_y})$, contributions from loops which encircle the
cylinder.  Keeping just the leading of these terms, we have the
slight modification of Eq.~\eqref{eq:4}
\begin{equation}
  \label{eq:31}
   H  =   -K \sum_{\Box}  \prod_{\langle ij\rangle \in \Box}
  \sigma_{ij}^z - t_s \sum_x \prod_{y=1}^{L_y} \sigma^z_{xy;xy+1}- h
  \sum_{\langle ij\rangle} \sigma_{ij}^x ,
\end{equation}
where we expect $t_s \sim e^{-L_y/\xi_y}$, which physically is
related to the amplitude for a virtual spinon to encircle the
cylinder.  Note that in this case $Q$ no longer commutes with $H$,
and the nature of the eigenstates is no longer clear.  Now if we
assume $t_s \ll K, h$ and that for $t_s=0$ we are in the deconfined
$Z_2$ phase, we can project the Hamiltonian on the low-energy sector
of the pure gauge theory, i.e. the two level system of the
quasi-degenerate states. Then we obtain the effective Hamiltonian,
written in a pseudo-spin notation in which $\mu^z=\mp 1$ correspond
to the vison/no-vison states:
\begin{equation}
  \label{eq:32}
  H_{deg} = - t_v \,\mu^x - t_s L_x \,\mu^z.
\end{equation}
Since $t_v \sim e^{-L_x/\xi_x}$ and $t_s \sim e^{-L_y/\xi_y}$, the
nature of the ground state depends crucially on the aspect ratio of
the cylinder.  For a fat cylinder, with small $L_x/L_y$, for which
$t_s \ll t_v$, the eigenstates will be like those of the pure gauge
theory, and the gap will be exponentially small in $L_x$.

However, for a ``long'' cylinder, with larger $L_x/L_y$, the gap
will be exponential instead in $L_y$.  Indeed, strictly in the limit
of large $L_x$ and $L_y$ fixed, the higher energy state can no
longer be regarded as quasi-degenerate: its energy, relative to the
ground state, grows linearly with $L_x$, and so other states with
local, non-topological excitations will have lower energy.  The
conditions for $t_s$ to dominate are much less restrictive than
this, however, requiring only $t_s L_x \gg t_v$, or $\exp(L_x/\xi_x
- L_y/\xi_y) \gg 1/L_x$.  In this limit, the ground state is an
approximate eigenstate of $\mu^z$, i.e. a state of definite vison
number.  Because of the quasi-one-dimensional nature of the DMRG
technique, in the most effective regime of this technique, this is
the expected form of the ground state.  Note again that this regime
is missed by the pure gauge theory and also the quantum dimer model.

The nature of the absolute ground state obtained by DMRG has
implications for the entanglement entropy.  As shown recently by
Zhang {\sl et al}\cite{2011arXiv1111.2342Z}, the topological
entanglement entropy for a cut with non-trivial topology actually
depends upon the choice of quasi-degenerate wavefunction.  The
cylindrical cut studied here is precisely such a cut.  The results
of Ref.\cite{2011arXiv1111.2342Z} imply that the topological
entanglement entropy reaches its maximum and universal value (of
$-\ln 2$) when the ground state is a vison eigenstate, and takes a
smaller (in magitude) value for other superpositions of states, {\sl
vanishing} for the case of a vison superposition, as is obtained in
the absence of spinons. Thus the result of our numerical study in
the main text, in which we found rough agreement with the $-\ln 2$
value for the topological entanglement entropy, in fact is evidence
for such a vison eigenstate in the numerics, consistent with the
predicted effects of virtual spin fluctuations.

\end{document}